\documentclass[aps,prd,preprintnumbers,groupedaddress,nofootinbib,amssymb,notitlepage,eqsecnum]{revtex4-1}
\usepackage{here}
\usepackage[dvipdfmx]{graphicx}
\usepackage{amsmath,amsthm,amssymb}
\usepackage{bm}
\usepackage{color}

\usepackage{amsfonts}
\usepackage{dcolumn}
\usepackage{hyperref}
\allowdisplaybreaks[1]
\usepackage{stackengine}

\newcommand{\be}{\begin{equation}}  
\newcommand{\ee}{\end{equation}}
\newcommand{\ba}{\begin{eqnarray}}
\newcommand{\ea}{\end{eqnarray}}

\newcommand{\rd}{{\rm d}}

\newcommand{\bem}{\begin{bmatrix}}
\newcommand{\eem}{\end{bmatrix}}
\newcommand{\Mpl}{M_{\rm Pl}}

\newcommand{\G}{\mathcal{G}}
\newcommand{\dphi}{\delta \phi}

\allowdisplaybreaks

\begin{document}

\preprint{WUCG-22-12}
\title{Cosmological stability in $f(\phi,{\cal G})$ gravity}

\author{Shinji Tsujikawa}

\affiliation{Department of Physics, Waseda University, 3-4-1 Okubo, 
Shinjuku, Tokyo 169-8555, Japan}

\begin{abstract}

In gravitational theories where a canonical scalar field $\phi$ with 
a potential $V(\phi)$ is coupled to a Gauss-Bonnet (GB) term ${\cal G}$ 
with the Lagrangian $f(\phi,{\cal G})$, we study the cosmological stability 
of  tensor and scalar perturbations in the presence of a perfect fluid.
We show that, in decelerating cosmological epochs with a positive 
tensor propagation speed squared, 
the existence of nonlinear functions of ${\cal G}$ in $f$ always induces 
Laplacian instability of a dynamical scalar perturbation associated 
with the GB term. 
This is also the case for $f({\cal G})$ gravity, where the presence 
of nonlinear GB functions $f({\cal G})$ 
is not allowed during the radiation- and matter-dominated epochs. 
A linearly coupled GB term with $\phi$ of the form 
$\xi (\phi){\cal G}$ can be consistent with all the stability conditions, provided that the scalar-GB coupling is subdominant to the 
background cosmological dynamics.

\end{abstract}

\date{\today}


\maketitle

\section{Introduction}
\label{introsec}

General Relativity (GR) is a fundamental theory of gravity whose validity 
has been probed in Solar System experiments \cite{Will:2014kxa} 
and submillimeter laboratory tests \cite{Hoyle:2000cv,Adelberger:2003zx}.
Despite the success of GR describing gravitational interactions
in the Solar System, there have been long-standing cosmological 
problems such as the origins of inflation, dark energy, and 
dark matter. To address these problems, one typically introduces 
additional degrees of freedom (DOFs) beyond 
those appearing in GR \cite{Copeland:2006wr,Silvestri:2009hh,Tsujikawa:2010zza,Tsujikawa:2010zza,Clifton:2011jh,Joyce:2014kja,Koyama:2015vza,Heisenberg:2018vsk}. One of such new DOFs is a canonical scalar field $\phi$ 
with a potential $V(\phi)$ \cite{Fujii:1982ms,Sato:1980yn,Kazanas:1980tx,Guth:1980zm,Linde:1981mu,Albrecht:1982wi,Linde:1983gd,Ratra:1987rm,Wetterich:1987fm,Chiba:1997ej,Ferreira:1997hj,Caldwell:1997ii}. 
If the scalar field evolves slowly along 
the potential, it is possible to realize cosmic acceleration 
responsible for inflation or dark energy. 
An oscillating scalar field around the potential minimum can 
be also the source for dark matter. 

The other way of introducing a new dynamical DOF is to modify the gravitational 
sector from GR. The Lagrangian in GR is given by an Einstein-Hilbert term 
$\Mpl^2 R/2$, where $\Mpl$ is the reduced Planck mass 
and $R$ is the Ricci scalar. 
If we consider theories containing nonlinear functions of $R$ of 
the form $f(R)$, there is one scalar DOF arising from 
the modification of gravity \cite{Sotiriou:2008rp,DeFelice:2010aj}.  
One well known example is the Starobinsky's model, in which 
the presence of a quadratic curvature term $R^2$ drives 
cosmic acceleration \cite{Starobinsky:1980te}. 
It is also possible to construct $f(R)$ models of late-time 
cosmic acceleration \cite{Capozziello:2002rd,Carroll:2003wy,Hu:2007nk,Amendola:2006we,Starobinsky:2007hu,Tsujikawa:2007xu,Linder:2009jz}, while being consistent with local gravity constraints.

The Einstein tensor $G_{\mu \nu}$ obtained by varying the Einstein-Hilbert 
action satisfies the conserved relation $\nabla^{\mu}G_{\mu \nu}=0$ 
($\nabla^{\mu}$ is a covariant derivative operator), 
with the property of second-order field equations of motion 
in metrics. If we demand such conserved and second-order properties for 
2-rank symmetric tensors, GR is the unique theory of gravity 
in 4 dimensions \cite{Lovelock:1971yv}. 
In spacetime dimensions higher than 4, there is 
a particular combination known as a Gauss-Bonnet (GB) term 
${\cal G}$ consistent with those 
demands \cite{Stelle:1977ry}. 
In 4 dimensions, the GB term is a topological surface term 
and hence it does not contribute to the field equations of motion.
In the presence of a coupling between a scalar field $\phi$ and $\G$ 
of the form $\xi(\phi)\G$, the spacetime dynamics is modified by 
the time or spatial variation of $\phi$.
Indeed, this type of scalar-GB coupling appears in the context of low energy effective 
string theory \cite{Zwiebach:1985uq,Gross:1986mw,Metsaev:1987zx}. 
The cosmological application of the coupling $\xi(\phi)\G$
has been extensively performed in the literature \cite{Antoniadis:1993jc,Gasperini:1996fu,Kawai:1998ab,Cartier:1999vk,Cartier:2001is,Tsujikawa:2002qc,Toporensky:2002ta,Amendola:2005cr,Nojiri:2005vv,Calcagni:2005im,Calcagni:2006ye,Koivisto:2006xf,Koivisto:2006ai,Tsujikawa:2006ph,Guo:2006ct,Amendola:2007ni,Satoh:2008ck,Guo:2009uk,Kanti:2015pda,Hikmawan:2015rze,Kawai:2017kqt,Yi:2018gse,Kawai:2021edk,Zhang:2021rqs,Kawai:2021bye,Kawaguchi:2022nku}. 
Moreover, it is known that the same coupling gives rise to spherically 
symmetric solutions of hairy black holes and neutron stars \cite{Kanti:1995vq,Torii:1996yi,Chen:2006ge,Guo:2008hf,Pani:2009wy,Kleihaus:2011tg,Sotiriou:2013qea,Ayzenberg:2014aka,Maselli:2015tta,Doneva:2017bvd,Silva:2017uqg,Antoniou:2017acq,Minamitsuji:2018xde,Silva:2018qhn,Langlois:2022eta,Minamitsuji:2022vbi,Minamitsuji:2022tze}.
The Lagrangian $f(\G)$ containing nonlinear 
functions of $\G$ also generates nontrivial contributions 
to the spacetime dynamics \cite{Carroll:2004de,Chiba:2005nz,Nojiri:2005jg,DeFelice:2006pg,Li:2007jm,DeFelice:2008wz,DeFelice:2009aj,Zhou:2009cy,Myrzakulov:2010gt,Bamba:2017cjr}.

In Ref.~\cite{DeFelice:2009ak}, De Felice and Suyama studied the stability of 
scalar perturbations in $f(R,\G)$ gravity on a spatially 
flat Friedmann-Lema\^{i}tre-Robertson-Walker (FLRW) background. 
In theories with $f_{,R \G}^2-f_{,RR}f_{,\G \G} \neq 0$, where 
$f_{,R \G}=\partial^2 f/\partial \G \partial R$, 
$f_{,RR}=\partial^2 f/\partial R^2$, and 
$f_{,\G \G}=\partial^2 f/\partial \G^2$, 
there is an unusual scale-dependent sound speed which propagates 
superluminally in the short-wavelength limit, unless the vacuum is 
in a de Sitter state (see also Ref.~\cite{DeFelice:2010hg} for 
the analysis in an anisotropic cosmological background). 
We note that this problem does not arise for 
$f(R)$ gravity or $\Mpl^2R/2+f(\G)$ gravity. 
In Ref.~\cite{DeFelice:2009wp}, the same authors extended the analysis 
to a more general Lagrangian $f(\phi,R,\G)$ with a canonical scalar 
field $\phi$ and showed that the property of the scale-dependent sound 
speed is not modified by the presence of $\phi$. 
Taking a perfect fluid (radiation or nonrelativistic matter) into account 
in $f(R,\G)$ gravity, the cosmological stability and evolution of matter 
perturbations were studied in 
Refs.~\cite{DeFelice:2010sh,DeFelice:2010hb,delaCruz-Dombriz:2011oii}. 

In Einstein-scalar-GB gravity given by the Lagrangian 
$\Mpl^2R/2+f(\phi,\G)$, where $\phi$ is a canonical scalar field,  
the problem of scale-dependent sound speeds mentioned above 
is not present. In this theory, the propagation of scalar perturbations 
on the spatially flat FLRW background was studied 
in Ref.~\cite{DeFelice:2009wp} without taking into account matter.
While the sound speed associated with the field $\phi$ is luminal for 
theories with $f_{,\G \G} \neq 0$, 
the propagation speed squared $c_s^2$ arising from a nonlinear GB term 
deviates from that of light and it can be even negative.
In Ref.~\cite{DeFelice:2009wp}, the authors discussed the possibility 
for satisfying the Laplacian stability condition $c_s^2>0$. 
In the presence of matter, however, the stability conditions are subject to 
modifications from those in the vacuum. To understand what happens 
for the dynamics of cosmological perturbations during radiation- and 
matter-dominated epochs, we need to study their stabilities 
by incorporating radiation or nonrelativistic matter.

In this letter, we will derive general conditions for the absence of ghosts and 
Laplacian instabilities in $\Mpl^2R/2+f(\phi,\G)$ gravity, where $\phi$ is 
a canonical scalar field with a potential $V(\phi)$. 
In theories where the scalar field $\phi$ is coupled to the linear GB term, 
i.e., $f(\phi,\G)=\xi(\phi)\G$, there is only one dynamical scalar DOF $\phi$.
In theories with $f_{,\G \G} \neq 0$, the Lagrangian $f(\phi,\G)$ can be 
expressed in terms of two scalar fields $\phi$ and $\chi$ coupled to 
the linear GB term, where $\chi$ arises from the nonlinearity in $\G$. 
Hence the latter theory has two dynamical scalar DOFs.
To study the cosmological stability of $f(\phi,\G)$ theories 
with $f_{,\G \G} \neq 0$, we take a perfect fluid into account 
as a form of the Schutz-Sorkin action \cite{Schutz:1977df,Brown:1992kc,DeFelice:2009bx}.
We will show that the squared sound speed arising from nonlinear functions of 
$\G$ is negative during decelerating cosmological epochs 
including radiation and matter eras. 
To reach this conclusion, we exploit the fact that the propagation 
speed squared $c_t^2$ of tensor perturbations must be positive
to avoid Laplacian instability of gravitational waves.

The same Laplacian instability of scalar perturbations is also present 
in $\Mpl^2R/2+f(\G)$ gravity with any nonlinear function 
of $\G$ in $f$. 
We note that, in $f(\G)$ models of late-time cosmic acceleration,  
violent instabilities of matter density perturbations during the 
radiation and matter eras were reported in Ref.~\cite{DeFelice:2009rw}. 
This can be regarded as the consequence of a negative sound speed squared 
of the scalar perturbation $\delta \G$ arising from the nonlinearity 
of $\G$ in $f$. Since $\delta \G$ is coupled to 
the matter perturbation $\delta \rho$, the background cosmological 
evolution during the radiation and matter eras is spoiled by 
the rapid growth of $\delta \rho$. 
Our analysis in this letter shows that similar catastrophic instabilities 
persist for more general scalar-GB couplings $f(\phi,\G)$ 
with $f_{,\G \G} \neq 0$.
 
This letter is organized as follows.
In Sec.~\ref{Fphisec}, we revisit cosmological stability conditions 
in $\Mpl^2R/2+\xi(\phi)\G$ gravity with a canonical scalar field $\phi$, 
which can be accommodated in a subclass of Horndeski 
theories with a single scalar 
DOF \cite{Horndeski,Def11,KYY,Charmousis:2011bf}.
This is an exceptional case satisfying the condition $f_{,\G \G}=0$,  
under which the Laplacian instability of scalar perturbations can be avoided.
In Sec.~\ref{backsec}, we derive the background equations and 
stability conditions of tensor perturbations in $\Mpl^2R/2+f(\phi,\G)$ 
gravity with $f_{,\G \G} \neq 0$ by incorporating a perfect fluid.
In Sec.~\ref{scalarsec}, we proceed to the derivation of 
a second-order action of scalar perturbations and 
obtain conditions for the absence of ghosts and 
Laplacian instabilities in the scalar sector. 
In particular, we show that an effective cosmological
equation of state $w_{\rm eff}$ needs to be in the range 
$w_{\rm eff}<-(2+c_t^2)/6$ to ensure 
Laplacian stabilities of the perturbation $\delta \G$.
Sec.~\ref{consec} is devoted to conclusions.

\section{$\xi(\phi)\G$ gravity}
\label{Fphisec}

We first briefly revisit the cosmological stability in 
$\xi(\phi)\G$ gravity given by the action 
\be
{\cal S}=\int {\rm d}^4 x \sqrt{-g} \left[ 
\frac{\Mpl^2}{2}R+ \eta X-V(\phi)
+\xi(\phi) \G \right]+{\cal S}_m (g_{\mu \nu}, \Psi_m)\,,
\label{action0}
\ee
where $g$ is a determinant of the metric tensor $g_{\mu \nu}$, 
$\eta$ is a constant, $X=-(1/2)g^{\mu \nu} \nabla_{\mu} \phi \nabla_{\nu} \phi$ 
is a kinetic term of the scalar field $\phi$, $V(\phi)$ and $\xi(\phi)$ are 
functions of $\phi$, and $\G$ is a GB term defined by 
\be
\G \equiv{R^{2}-4R_{\mu\nu}R^{\mu\nu}
+R_{\mu\nu\rho\sigma}R^{\mu\nu\rho\sigma}}\,.
\ee
with $R_{\mu\nu}$ and $R_{\mu\nu\rho\sigma}$ being 
the Ricci and Riemann tensors, respectively. 
For the matter action ${\cal S}_m$, we consider 
a perfect fluid minimally coupled to gravity. 

The action (\ref{action0}) contains one scalar DOF 
$\phi$ besides the matter field $\Psi_m$. 
If we consider Horndeski theories \cite{Horndeski} 
given by the action
\ba
{\cal S}
&=& 
\int {\rm d}^4 x \sqrt{-g}\,
\biggl[ G_2(\phi,X)-G_{3}(\phi,X)\square\phi 
+G_{4}(\phi,X)R +G_{4,X}(\phi,X)
\left\{ (\square \phi)^{2}
-(\nabla_{\mu}\nabla_{\nu} \phi)
(\nabla^{\mu}\nabla^{\nu} \phi) \right\} \nonumber \\
& &
+G_{5}(\phi,X)G_{\mu \nu} \nabla^{\mu}\nabla^{\nu} \phi
-\frac{1}{6}G_{5,X}(\phi,X)
\left\{ (\square \phi )^{3}-3(\square \phi)\,
(\nabla_{\mu}\nabla_{\nu} \phi)
(\nabla^{\mu}\nabla^{\nu} \phi)
+2(\nabla^{\mu}\nabla_{\alpha} \phi)
(\nabla^{\alpha}\nabla_{\beta} \phi)
(\nabla^{\beta}\nabla_{\mu} \phi) \right\} 
\biggr] 
\nonumber \\
& &
+{\cal S}_m (g_{\mu \nu}, \Psi_m)\,,
\label{actionHo}
\ea
then the theory (\ref{action0}) can be accommodated by 
choosing the coupling functions \cite{KYY}
\ba
& &
G_2(\phi,X)=\eta X-V(\phi)
+8 \xi_{,\phi \phi \phi \phi}(\phi) X^2 (3-\ln |X|)\,,\qquad 
G_3(\phi,X)=4 \xi_{,\phi \phi \phi}(\phi) X (7-3\ln |X|)\,,\nonumber \\
& &
G_4(\phi,X)=\frac{\Mpl^2}{2}+4 \xi_{,\phi \phi}(\phi) X (2-\ln |X|)\,,\qquad
G_5(\phi,X)=-4 \xi_{,\phi}(\phi) \ln |X|\,,
\label{Ggauss}
\ea
where we use the notations 
$F_{,X}=\partial F/\partial X$ and 
$F_{,\phi}=\partial F/\partial \phi$ for any arbitrary function $F$. 

Let us consider a spatially flat FLRW background given by 
the line element $\rd s^2=-\rd t^2+a^2(t) \delta_{ij} \rd x^i \rd x^j$, 
where $a(t)$ is a time-dependent scale factor. 
The perfect fluid has a density $\rho$ and 
pressure $P$. The background equations as well as the perturbation 
equations in full Horndeski theories were 
derived in Refs.~\cite{KYY,DeFelice:2011hq,DeFelice:2011bh,Kase:2018aps}.
On using the correspondence (\ref{Ggauss}), 
the background equations of motion in theories 
given by the action (\ref{action0}) are 
\ba
& & 3 \tilde{q}_t H^2=\frac{1}{2} \eta \dot{\phi}^2+V(\phi)+\rho\,,
\label{back0a} \\
& & 2 \tilde{q}_t \dot{H}=-\eta \dot{\phi}^2-H^2  \tilde{q}_t 
\left( \tilde{c}_t^2-1 \right)-\rho-P\,,
\label{back1a} \\
& & \eta \left( \ddot{\phi}+3H \dot{\phi} \right)
+V_{,\phi}-\xi_{,\phi}\G=0\,,
\label{back2a}\\
& & \dot{\rho}+3H (\rho+P)=0\,,
\ea
where $H=\dot{a}/a$ is the Hubble expansion rate, a dot represents 
the derivative with respect to $t$, and 
\ba
\tilde{q}_t &=& \Mpl^2+8 \xi_{,\phi} H \dot{\phi}\,,
\label{qt0} \\
\tilde{c}_t^2 &=& \frac{\Mpl^2+8 (\xi_{,\phi} \ddot{\phi}
+\xi_{,\phi \phi}\dot{\phi}^2)}
{\Mpl^2+8 \xi_{,\phi} H \dot{\phi}}\,,\label{ct0} \\
\G &=& 24H^2 \left( H^2+\dot{H} \right)\,.
\ea

In the presence of tensor perturbations $h_{ij}$ with the perturbed 
line element  $\rd s^2=-\rd t^2+a^2(t) (\delta_{ij}+h_{ij}) \rd x^i \rd x^j$,
the second-order action of traceless and divergence-free modes of 
$h_{ij}$ was already derived in full Horndeski theories \cite{KYY,DeFelice:2011bh,Kase:2018aps}. 
In the current theory, the conditions for the absence of ghosts 
and Laplacian instabilities are 
\ba
& &
\tilde{q}_t>0\,,\label{qt1}\\
& &
\tilde{c}_t^2>0\,,\label{ct1}
\ea
where $\tilde{q}_t$ and $\tilde{c}_t^2$ are defined by Eqs.~(\ref{qt0}) 
and (\ref{ct0}), respectively. 
Note that $\tilde{q}_t$ determines the sign of a kinetic term of $h_{ij}$, 
while $\tilde{c}_t^2$ corresponds to the propagation speed squared 
of tensor perturbations. 

For the scalar sector, we choose the perturbed line element 
$\rd s^2=-(1+2\alpha) \rd t^2+2 \partial_i B \rd t \rd x^i
+a^2(t)\delta_{ij}\rd x^i \rd x^j$ in the flat gauge, where 
$\alpha$ and $B$ are scalar metric perturbations. 
There is also a scalar-field perturbation $\delta \phi$ besides 
the matter perturbation $\delta \rho$ and the 
fluid velocity potential $v$. 
After deriving the quadratic-order action of scalar perturbations, 
we can eliminate nondynamical variables $\alpha$, $B$, 
and $v$ from the action. 
Then, we are left with the two dynamical perturbations 
$\delta \phi$ and $\delta \rho$ in the second-order action. 
In the short-wavelength limit, there is neither ghost nor 
Laplacian instability for $\delta \phi$ under 
the conditions \cite{KYY,DeFelice:2011bh,Kase:2018aps}
\ba
\tilde{q}_s &=& 2 \left( \eta \tilde{q}_t
+96 H^4 \xi_{,\phi}^2 \right)>0\,,
\label{qs0} \\
\tilde{c}_s^2 &=& \frac{\eta \tilde{q}_t-32 (2+\tilde{c}_t^2+6w_{\rm eff})
H^4 \xi_{,\phi}^2}{\eta \tilde{q}_t+96 H^4 \xi_{,\phi}^2}>0\,,
\label{cs0}
\ea
where $\tilde{c}_s$ corresponds to the propagation speed 
of $\delta \phi$, and $w_{\rm eff}$ is the cosmological 
effective equation of state defined by 
\be
w_{\rm eff} \equiv -1-\frac{2\dot{H}}{3H^2}\,.
\label{weff}
\ee
The stability conditions for $\delta \rho$ are 
given by $\rho+P>0$ and $c_m^2>0$, where 
$c_m^2$ is the matter sound speed squared.

Under the stability condition (\ref{qt1}) with $\eta>0$, 
the scalar no-ghost condition (\ref{qs0}) is satisfied.  
Let us consider the case in which contributions of 
the scalar-GB coupling are suppressed, such that 
\be
\{ |\xi_{,\phi}H \dot{\phi}|, 
|\xi_{,\phi}\ddot{\phi}|, 
|\xi_{,\phi \phi} \dot{\phi}^2| \} \ll \Mpl^2\,,\qquad 
H^4 \xi_{,\phi}^2 \ll \eta \tilde{q}_t\,.
\label{GBlimit}
\ee
Then, it follows that $\tilde{q}_t \simeq \Mpl^2$, $\tilde{c}_t^2 \simeq 1$, 
$\tilde{q}_s \simeq 2\eta \Mpl^2$, and $\tilde{c}_s^2 \simeq 1$. 
In such cases, provided that $\eta>0$, all the stability conditions 
are consistently satisfied. 
If the scalar-GB coupling contributes to the late-time 
cosmological dynamics, there is an observational bound on $\tilde{c}_t$
constrained from the GW170817 event together with 
the electromagnetic counterpart, i.e., 
$-3 \times 10^{-15} \le \tilde{c}_t-1 \le 7 \times 10^{-16}$ 
\cite{LIGOScientific:2017zic} for the redshift $z \le 0.009$. 
This translates to the limit 
\be
\left| \xi_{,\phi} \ddot{\phi}+\xi_{,\phi \phi} \dot{\phi}^2
-\xi_{,\phi} H \dot{\phi} \right| \lesssim 10^{-15} \Mpl^2\,,
\label{ctbound}
\ee
which gives a tight constraint on the amplitude of $\xi(\phi)$.
In this case, contributions of the scalar-GB coupling to the background 
Eqs.~(\ref{back0a}) and (\ref{back1a}) are highly suppressed 
relative to the field density $\rho_{\phi}=\eta \dot{\phi}^2/2
+V(\phi)$ and the matter density.

The bound (\ref{ctbound}) is not applied to early cosmological epochs 
including inflation, radiation, and matter eras. 
We note, however, that the dominance of the scalar-GB coupling 
to the background equations prevents the successful 
cosmic expansion history. This can also give rise to the 
violation of either of the stability conditions (\ref{qt1})-(\ref{cs0}).
Provided the scalar-GB coupling is suppressed in such 
a way that inequalities (\ref{GBlimit}) hold, the linear stabilities 
are ensured for both tensor and scalar perturbations. 

\section{$f(\phi,\G)$ gravity}
\label{backsec}

We extend $\xi(\phi)\G$ gravity to more general theories 
in which a canonical scalar field $\phi$ with a potential $V(\phi)$ 
is coupled to the GB term of the form $f(\phi,\G)$. 
The action in such theories is given by 
\be
{\cal S}=\int {\rm d}^4 x \sqrt{-g} \left[ 
\frac{\Mpl^2}{2}R+ \eta X-V(\phi)
+f(\phi,\G) \right]+{\cal S}_m (g_{\mu \nu}, \Psi_m)\,,
\label{action}
\ee
where a matter field $\Psi_m$ is minimally coupled to gravity. 
It is more practical to introduce a scalar field $\chi$ and 
resort to the following action 
\be
{\cal S}=\int {\rm d}^4 x \sqrt{-g} \left[ 
\frac{\Mpl^2}{2}R+ \eta X-U(\phi,\chi)
+\xi(\phi,\chi)\G \right]+{\cal S}_m (g_{\mu \nu}, \Psi_m)\,,
\label{action2}
\ee
where 
\be
U(\phi,\chi) \equiv V(\phi)-f(\phi,\chi)
+\chi \xi(\phi,\chi)\,,\qquad 
\xi(\phi, \chi) \equiv f_{,\chi}(\phi,\chi)\,,
\ee
with the notation $f_{,\chi}=\partial f/\partial \chi$.
Varying the action (\ref{action2}) with respect to $\chi$, 
it follows that 
\be
\left( \chi-\G \right) \xi_{,\chi}=0\,.
\label{chiGre}
\ee
So long as $\xi_{,\chi} \neq 0$, we obtain $\chi=\G$. 
In this case, the action (\ref{action2}) reduces to (\ref{action}). 
Thus, the equivalence of (\ref{action2}) with (\ref{action}) holds for 
\be
f_{,\G \G} \neq 0\,,
\label{fcon}
\ee
under which there is a new scalar DOF $\chi$ arising from 
the gravitational sector.

Theories with $f_{,\G \G}=0$ correspond to the coupling $f=\xi(\phi)\G$, 
in which case the cosmological stability conditions were already 
discussed in Sec.~\ref{Fphisec}. 
In $\xi(\phi)\G$ gravity, we do not have the additional scalar DOF $\chi$ 
arising from $\G$, so the term $\xi_{,\chi}$ in Eq.~(\ref{chiGre}) 
does not have the meaning of $f_{,\G \G}$. 
Thus, the action (\ref{action2}) with the new dynamical DOF $\chi$ does not 
reproduce the action (\ref{action0}) in $\xi(\phi)\G$ gravity.

In the following, we will focus on theories with $f_{,\G \G} \neq 0$, i.e., 
those containing the nonlinear dependence of $\G$ in $f$.
For the matter field $\Psi_m$, we incorporate a perfect fluid 
without a dynamical vector DOF.
This matter sector is described by the Schutz-Sorkin 
action \cite{Schutz:1977df,Brown:1992kc,DeFelice:2009bx}
\be
{\cal S}_m=-\int {\rm d}^4 x \left[ \sqrt{-g}\,\rho (n)
+J^{\mu} \partial_{\mu}\ell \right]\,,
\label{Sm}
\ee
where the fluid density $\rho$ is a function of its number density $n$. 
The vector field $J^{\mu}$ is related to $n$ according to 
the relation $n=\sqrt{J^{\mu}J^{\nu}g_{\mu \nu}/g}$, 
where $u^{\mu}=J^{\mu}/(n \sqrt{-g})$ is the fluid four velocity. 
A scalar quantity $\ell$ in ${\cal S}_m$ is a Lagrange multiplier,  
with the notation of a partial derivative 
$\partial_{\mu}\ell=\partial \ell/\partial x^{\mu}$. 
Varying the matter action (\ref{Sm}) with respect to $\ell$ 
and $J^{\mu}$, respectively, we obtain
\ba
& &
\partial_{\mu} J^{\mu}=0\,,
\label{Jmucon} \\
& &
\partial_{\mu}\ell=u_{\mu} \rho_{,n}\,,
\label{leq} 
\ea
where $\rho_{,n}=\rd \rho/\rd n$.

\subsection{Background equations}

We derive the background equations of motion on the 
spatially flat FLRW
background given by the line element
\be
{\rm d}s^2=-N^2(t)\,\rd t^2+a^2(t) \delta_{ij} \rd x^i \rd x^j\,,
\label{element}
\ee
where $N(t)$ is a lapse function. 
Since the fluid four velocity in its rest frame is given by 
$u^{\mu}=(N^{-1}, 0, 0, 0)$, the vector field $J^{\mu}$ 
has components $J^{\mu}=(n a^3,0,0,0)$. 
{}From Eq.~(\ref{Jmucon}), we obtain
\be
{\cal N}_0 \equiv n a^3={\rm constant}\,,
\ee
which means that the total fluid number ${\cal N}_0$ 
is conserved. This translates to the differential equation 
$\dot{n}+3H n=0$, which can be expressed as a form 
of the continuity equation 
\be
\dot{\rho}+3H \left( \rho+P \right)=0\,,
\label{contieq}
\ee
where $P$ is a fluid pressure defined by $P=n \rho_{,n}-\rho$.

On the background (\ref{element}), the total action (\ref{action2}) 
is expressed in the form 
\be
{\cal S}=\int \rd t\, \rd^3 x  \left[ \frac{\eta a^3 \dot{\phi}^2}{2N}
-\frac{3\Mpl^2 a \dot{a}^2}{N}-N a^3 U(\phi,\chi)
-\frac{8 \dot{a}^3 \dot{\xi}(\phi,\chi)}{N^3}
-N a^3 \rho-n a^3 \dot{\ell} \right]\,.
\label{action3}
\ee
{}From Eq.~(\ref{leq}), we obtain the following relation 
\be
\dot{\ell}=-N \rho_{,n}\,.
\ee
Varying the action (\ref{action3}) with respect to $N$, $a$, 
$\phi$, $\chi$ respectively and setting $N=1$ at the end, we obtain 
the background equations of motion
\ba
& &
3 q_t H^2=\frac{1}{2} \eta \dot{\phi}^2
+U(\phi,\chi)+\rho\,,
\label{back1} \\
& &
2 q_t \dot{H}=-\eta \dot{\phi}^2-H^2 q_t \left( c_t^2-1 \right)
-\rho-P\,,
\label{back2} \\
& &
\eta \left( \ddot{\phi}+3H \dot{\phi} \right)
+V_{,\phi}-f_{,\phi}=0\,,
\label{back3}\\
& &
\chi=\G=24H^2 \left( H^2+\dot{H} \right)\,,
\ea
where 
\ba
q_t &=& \Mpl^2+8 H ( \xi_{,\phi} \dot{\phi}
+\xi_{,\chi} \dot{\chi} )\,,\label{qt}\\
c_t^2 &=& \frac{\Mpl^2+8 
(\xi_{,\phi}\ddot{\phi}
+\xi_{,\phi \phi}\dot{\phi}^2
+\xi_{,\chi} \ddot{\chi}
+\xi_{,\chi \chi} \dot{\chi}^2
+2\xi_{,\phi \chi} \dot{\phi} \dot{\chi})}
{\Mpl^2+8 H (\xi_{,\phi} \dot{\phi}
+\xi_{,\chi} \dot{\chi})}\,.
\label{ct}
\ea

We recall that the perfect fluid obeys the continuity 
Eq.~(\ref{contieq}). 
We notice that Eqs.~(\ref{back1})-(\ref{back3}) are of similar forms 
to Eqs.~(\ref{back0a})-(\ref{back2a}) in $\xi(\phi)\G$ gravity, 
but the expressions of $q_t$ and $c_t^2$ are different from 
$\tilde{q}_t$ and $\tilde{c}_t^2$, respectively, 
because of the appearance of $\chi$-dependent terms. 
These new terms do not vanish for $\xi_{,\chi} \neq 0$, 
i.e., for $f_{,\G \G} \neq 0$. 
As we will show in Sec.~\ref{scalarsec}, nonlinearities 
of $\G$ in $f$ are responsible for the appearance of 
a new scalar propagating DOF $\delta \chi$.

\subsection{Stabilities in the tensor sector}
\label{tensorsec}

We proceed to the derivation of stability conditions for 
tensor perturbations in theories given by the action (\ref{action2}). 
The perturbed line element including the tensor perturbation 
$h_{ij}$ is 
\be
\rd s^2=-\rd t^2+a^2(t) \left( \delta_{ij}+h_{ij} \right) 
\rd x^i \rd x^j\,,
\ee
where we impose the traceless and divergence-free gauge conditions 
${h^{i}}_{i}=0$ and $\partial^{i}h_{ij}=0$.
For the gravitational wave propagating along the $z$ direction, 
nonvanishing components of $h_{ij}$ are 
expressed in the form  
\be
h_{11}=h_1(t,z)\,,\qquad 
h_{22}=-h_1(t,z)\,,\qquad
h_{12}=h_{21}=h_2(t,z)\,,
\ee
where the two polarized modes $h_1$ and $h_2$ are functions 
of $t$ and $z$.

The second-order action arising from the matter action 
(\ref{Sm}) can be expressed as
\be
({\cal S}_m^{(2)})_t=-\int \rd t\,\rd^3 x \sum_{i=1}^{2} 
\frac{1}{2}a^3 P h_i^2\,,
\ee
where $P$ can be eliminated by using the background 
Eq.~(\ref{back2}). 
Expanding the total action (\ref{action2}) up to quadratic 
order in tensor perturbations and integrating it by 
parts, the resulting second-order action reduces to
\be
{\cal S}_t^{(2)}=\int \rd t\,\rd^3x \sum_{i=1}^{2}
\frac{a^3}{4}q_t \left[ \dot{h}_i^2-\frac{c_t^2}{a^2} 
(\partial h_i)^2 \right]\,,
\label{actionSt}
\ee
where $(\partial h_i)^2=(\partial h_i/\partial z)^2$. 
We recall that $q_t$ and $c_t^2$ are given by 
Eqs.~(\ref{qt}) and (\ref{ct}), respectively.

To avoid the ghost and Laplacian instabilities 
in the tensor sector, we require the two conditions 
$q_t>0$ and $c_t^2>0$, which translate to 
\ba
& &
\Mpl^2+8 H ( \xi_{,\phi} \dot{\phi}
+\xi_{,\chi} \dot{\chi} )>0\,,\label{qtcon}\\
& &
\Mpl^2+8 
(\xi_{,\phi}\ddot{\phi}
+\xi_{,\phi \phi}\dot{\phi}^2
+\xi_{,\chi} \ddot{\chi}
+\xi_{,\chi \chi} \dot{\chi}^2
+2\xi_{,\phi \chi} \dot{\phi} \dot{\chi})>0\,.
\label{ctcon}
\ea
In $f(\G)$ gravity without the scalar field $\phi$, 
tensor stability conditions can be obtained by 
setting $\dot{\phi}=0$ and $\ddot{\phi}=0$ in Eqs.~(\ref{qtcon}) 
and (\ref{ctcon}).

We vary the action (\ref{actionSt}) with respect to $h_i$ (with $i=1,2$) 
in Fourier space with a comoving wavenumber ${\bm k}$.
Then, we obtain the tensor perturbation equation of motion 
\be
\ddot{h}_i+\left(3H +\frac{\dot{q}_t}{q_t} \right) \dot{h}_i
+c_t^2 \frac{k^2}{a^2} h_i=0\,,
\label{heq}
\ee
where $k=|{\bm k}|$.
Since $\xi=f_{,\chi}=f_{,\G}$, the $\G$ dependence in $f$ 
leads to the modified evolution of gravitational waves 
in comparison to GR. 
If the energy densities of $\phi$ and $\chi$ are relevant to 
the late-time cosmological dynamics after the matter dominance, 
the observational constraint on the tensor propagation speed $c_t$ 
arising from the GW170817 event \cite{LIGOScientific:2017zic} 
($|c_t-1| \lesssim 10^{-15}$) 
gives a tight bound on the scalar-GB coupling $f(\phi,\G)$. 
Such a stringent limit is not applied to the cosmological dynamics 
in the early Universe, but the conditions (\ref{qtcon}) and 
(\ref{ctcon}) need to be still satisfied.

\section{Stabilities of $f(\phi,\G)$ gravity in the scalar sector}
\label{scalarsec}

In this section, we will derive conditions for the absence of 
scalar ghosts and Laplacian instabilities in theories given by 
the action (\ref{action2}).
A perturbed line element containing scalar perturbations 
$\alpha$, $B$, $\zeta$, and $E$ is of the form
\be
\rd s^2=-(1+2\alpha) \rd t^2+2 \partial_i B \rd t \rd x^i
+a^2(t) \left[ (1+2\zeta) \delta_{ij}
+2\partial_i \partial_j E \right] 
\rd x^i \rd x^j\,.
\label{permet}
\ee
For the scalar fields $\phi$ and $\chi$, we consider 
perturbations $\delta \phi$ and $\delta \chi$ 
on the background values $\bar{\phi}(t)$ and 
$\bar{\chi}(t)$, respectively, such that 
\be
\phi=\bar{\phi}(t)+\delta \phi (t,{\bm x})\,,\qquad 
\chi=\bar{\chi}(t)+\delta \chi (t,{\bm x})\,,
\ee
where we will omit a bar from the background quantities 
in the following.

In the matter sector, the temporal and spatial components of 
$J^{\mu}$ are decomposed into the background 
and perturbed parts as
\be
J^{0} = \mathcal{N}_{0}+\delta J\,,\qquad
J^{i} =\frac{1}{a^2(t)}\,\delta^{ik}\partial_{k}\delta j\,,
\label{elldef}
\ee
where $\delta J$ and $\delta j$ are scalar perturbations. 
In terms of the velocity potential $v$, the spatial 
component of fluid four velocity is expressed as 
$u_i=-\partial_i v$. 
From Eq.~(\ref{leq}), the scalar quantity $\ell$ 
has a background part obeying the relation 
$\dot{\ell}=-\rho_{,n}$ besides a perturbation $-\rho_{,n}v$. 
Then, we have
\be
\ell=-\int^{t} \rho_{,n} 
(\tilde{t})\,\rd \tilde{t} 
-\rho_{,n}v\,.
\label{ells}
\ee
Defining the matter density perturbation 
\be
\delta \rho \equiv \frac{\rho_{,n}}{a^3} \left[\delta J
-\mathcal{N}_{0}(3\zeta+\partial^2E)\right]\,,
\label{drhom}
\ee
the fluid number density $n$ 
has a perturbation \cite{Heisenberg:2018wye,Kase:2018aps}
\be
\delta n= \frac{\delta \rho}{\rho_{,n}}
-\frac{({\cal N}_0 \partial \chi+\partial \delta j)^2}{2{\cal N}_0a^5}
-\frac{(3\zeta+\partial^2E)\delta \rho}{\rho_{,n}}
-\frac{{\cal N}_0(\zeta+\partial^2E)(3\zeta-\partial^2E)}{2a^3}\,,
\ee
up to second order. 
The matter sound speed squared is given by 
\be
c_m^2=\frac{P_{,n}}{\rho_{,n}}=
\frac{n \rho_{,nn}}{\rho_{,n}}\,.
\ee
Expanding (\ref{Sm}) up to quadratic order in perturbations, 
we obtain the second-order matter action same as that
derived in Refs.~\cite{Heisenberg:2018wye,Kase:2018aps}. 
Varying this matter action with respect to $\delta j$ leads to 
\be
\partial \delta j=-a^3 n \left( \partial v+\partial B 
\right)\,,
\ee
whose relation will be used to eliminate $\delta j$.

In the following, we choose the gauge 
\be
E=0\,,
\ee
under which a scalar quantity $\xi$ associated with the
spatial gauge transformation 
$x^{i} \to x^{i}+\delta^{ij} \partial_j \xi$ is fixed. 
A scalar quantity $\xi^0$ associated with the
temporal part of the gauge transformation 
$t \to t+\xi^{0}$ can be fixed 
by choosing a flat gauge ($\zeta=0$) or a unitary 
gauge ($\delta \phi=0$). 
We do not specify the temporal gauge condition in 
deriving the second-order action, 
but we will do so at the end.

Expanding the total action (\ref{action2}) up to quadratic order 
in scalar perturbations and integrating it by parts, 
the resulting second-order action is given by 
\be
{\cal S}_s^{(2)}=\int \rd t\,\rd^3x 
\left( L_{\rm flat}+L_{\zeta} \right)\,, 
\label{Ss}
\ee
where 
\ba
\hspace{-1.0cm}
L_{\rm flat}
&=& a^3 \biggl[
\frac{\eta}{2} \dot{\dphi}^2-\frac{\eta}{2}\frac{(\partial\dphi)^2}{a^2}+
\frac{1}{2} \left( f_{,\phi \phi}-V_{,\phi \phi} \right) \dphi^2
+\left\{ \frac{1}{2} \eta \dot{\phi}^2-3H^2 (2 q_t-\Mpl^2) \right\} \alpha^2
-H (3q_t-\Mpl^2) \frac{\partial^2 B}{a^2} \alpha \nonumber \\
& &~~~-\frac{C_4}{16 H^2} \delta \chi^2
+\biggl\{ C_1 \dot{\dphi}+C_2 \dphi-C_3 \frac{\partial^2\dphi}{a^2}
+3H C_4 \dot{\delta \chi}-C_4 \frac{\partial^2 \delta \chi}{a^2} 
+3(H C_5-\dot{H} C_4)\delta \chi \biggr\} \alpha \nonumber \\
& &~~~
+\left( C_3 \dot{\delta \phi}+C_6 \delta \phi
+C_4 \dot{\delta \chi}+C_5 \delta \chi \right) \frac{\partial^2 B}{a^2}
+\left( \rho+P \right)v\frac{\partial^2 B}
{a^2}-v\dot{\delta \rho}-3H (1+c_m^2) v\delta \rho  \nonumber \\
& &~~~
-\frac{1}{2} (\rho+P) \frac{(\partial v)^2}{a^2}
-\frac{c_m^2}{2 (\rho+P)} \delta \rho^2 
-\alpha \delta \rho \biggr]\,,
\label{LF}\\
\hspace{-1.0cm}
L_{\zeta}
&=& a^3 \biggl[ \biggl\{ 3H (3q_t-\Mpl^2)\alpha
-3 \left( C_3 \dot{\delta \phi}+C_6 \delta \phi
+C_4 \dot{\delta \chi}+C_5 \delta \chi \right)
-3(\rho+P)v+2q_t \frac{\partial^2 B}{a^2}
\biggr\} \dot{\zeta}-3q_t \dot{\zeta}^2 \nonumber \\
\hspace{-1.0cm}
& &~~~+q_t c_t^2 \frac{(\partial \zeta)^2}{a^2}
-2\biggl\{ q_t \alpha+\biggl( 1+\frac{\dot{H}}{H^2} 
\biggr) (C_3 \delta \phi+C_4 \delta \chi) \biggr\} 
\frac{\partial^2 \zeta}{a^2} \biggr]\,,
\label{Lze}
\ea
where $q_t$ and $c_t^2$ are given by 
Eqs.~(\ref{qt}) and (\ref{ct}), respectively, and 
\ba
& &
C_1=24 H^3 \xi_{,\phi}-\eta \dot{\phi}\,,\qquad 
C_2=-24H^2 \left[ (H^2+\dot{H}) \xi_{,\phi}
-H ( \xi_{,\phi \phi} \dot{\phi}+
\xi_{,\phi \chi} \dot{\chi}) \right]
-V_{,\phi}+f_{,\phi}\,,\nonumber \\
& &
C_3=8 H^2 \xi_{,\phi}\,,\qquad 
C_4=8 H^2 \xi_{,\chi}\,,\nonumber \\
& &
C_5=-8H^2 \left( H \xi_{,\chi}
-\xi_{,\phi \chi} \dot{\phi}
-\xi_{,\chi \chi} \dot{\chi} \right)\,,\qquad 
C_6=\eta \dot{\phi}-8 H^2 \left( 
H \xi_{,\phi}-\xi_{,\phi \phi} \dot{\phi}
-\xi_{,\phi \chi} \dot{\chi} \right)\,.
\ea
Now, we switch to the Fourier space with a comoving 
wavenumber ${\bm k}$. 
Varying the total action (\ref{Ss}) with respect to 
$\alpha$, $B$, and $v$, respectively, we obtain 
\ba
& &
C_1 \dot{\delta \phi}+C_2 \delta \phi+3H C_4 \dot{\delta \chi}
-3 (\dot{H}C_4-H C_5)\delta \chi
+3 \left( 3q_t-\Mpl^2 \right) H \dot{\zeta}
+\left[ \eta \dot{\phi}^2-6H^2 (2q_t-\Mpl^2) \right]\alpha 
\nonumber \\
& & 
+\frac{k^2}{a^2} \left[ 2q_t \zeta+H \left( 3q_t-\Mpl^2 \right)B
+C_3 \delta \phi+C_4 \delta \chi \right]-\delta \rho=0\,,
\label{pereq1} \\ 
& &
C_3 \dot{\delta \phi}+C_6 \delta \phi+2q_t \dot{\zeta}
+C_4 \dot{\delta \chi}+C_5 \delta \chi-H \left( 3q_t-\Mpl^2 
\right) \alpha+(\rho+P)v=0\,,
\label{pereq2} \\
& & 
\dot{\delta \rho}+3H (1+c_m^2) \delta \rho+3(\rho+P) \dot{\zeta}
+\frac{k^2}{a^2}(\rho+P) (v+B)=0\,.
\label{pereq3}
\ea

In the following, we choose the flat gauge given by 
\be
\zeta=0\,.
\label{flat}
\ee
to obtain stability conditions for scalar perturbations.
We will discuss the two cases: (A) $f(\phi,\G)$ gravity 
and (B) $f(\G)$ gravity in turn.

\subsection{$f(\phi,\G)$ gravity}

In $f(\phi,\G)$ gravity with $f_{,\G \G} \neq 0$, we can 
construct gauge-invariant scalar perturbations
$\delta \phi_{\rm f}=\delta \phi-\dot{\phi}\,\zeta/H$,  
$\delta \chi_{\rm f}=\delta \chi-\dot{\chi}\,\zeta/H$, and 
$\delta \rho_{\rm f}=\delta \rho-\dot{\rho}\,\zeta/H$. 
For the gauge choice (\ref{flat}), they reduce, respectively, 
to $\delta \phi$, $\delta \chi$, and $\delta \rho$, 
which correspond to the dynamical scalar DOFs. 
Note that the perturbation $\delta \chi=\delta \G$ arises from 
nonlinearities in the GB term. 
We solve Eqs.~(\ref{pereq1})-(\ref{pereq3}) for $\alpha$, 
$B$, $v$ and substitute them into Eq.~(\ref{Ss}). 
Then, the resulting quadratic-order action in Fourier space 
is expressed in the form 
\be
{\cal S}_s^{(2)}=\int \rd t\,\rd^3x\,a^{3}\left( 
\dot{\vec{\mathcal{X}}}^{t}{\bm K}\dot{\vec{\mathcal{X}}}
-\frac{k^2}{a^2}\vec{\mathcal{X}}^{t}{\bm G}\vec{\mathcal{X}}
-\vec{\mathcal{X}}^{t}{\bm M}\vec{\mathcal{X}}
-\vec{\mathcal{X}}^{t}{\bm B}\dot{\vec{\mathcal{X}}}
\right)\,,
\label{Ss2}
\ee
where ${\bm K}$, ${\bm G}$, ${\bm M}$, ${\bm B}$ are 
$3 \times 3$ matrices, and 
\be
\vec{\mathcal{X}}^{t}=\left(\delta \phi, \delta \chi, 
\delta \rho/k \right) \,.
\label{chit}
\ee
The leading-order contributions to 
${\bm M}$ and ${\bm B}$ are of order $k^0$. 
Taking the small-scale limit $k \to \infty$, nonvanishing 
components of the symmetric matrices 
${\bm K}$ and ${\bm G}$ are
\ba
& &
K_{11}=\frac{\eta [C_3 \dot{\phi}-H(3q_t-\Mpl^2)]^2
+6C_3^2 H^2 q_t}
{2H^2 (3q_t-\Mpl^2)^2}\,,\qquad 
K_{22}=\frac{C_4^2 (\eta \dot{\phi}^2+6H^2 q_t)}
{2H^2 (3q_t-\Mpl^2)^2}\,,\nonumber\\
& &
K_{12}=K_{21}=\frac{C_4
[C_3 ( \eta \dot{\phi}^2+6 H^2 q_t )
-\eta H \dot{\phi} (3q_t-\Mpl^2)]}{2H^2 (3q_t-\Mpl^2)^2}\,,
\qquad K_{33}=\frac{a^2}{2(\rho+P)}\,,
\ea
and 
\ba
& &
G_{11}=-\frac{\eta H(3q_t-\Mpl^2) 
[2C_3 \dot{\phi}-H(3q_t-\Mpl^2)]+C_3^2[\rho+P
-6 q_t \dot{H}+3H^2 q_t (c_t^2-3)]}
{2H^2 (3q_t-\Mpl^2)^2}\,, \nonumber\\
& &
G_{22}=\frac{C_4^2[ 3H^2 q_t (3-c_t^2)
+6 q_t \dot{H}-\rho-P]}
{2H^2 (3q_t-\Mpl^2)^2}\,,\nonumber\\
& &
G_{12}=G_{21}=-\frac{C_4[\eta H \dot{\phi} (3q_t-\Mpl^2)
+C_3 \{ \rho+P
-6 q_t \dot{H}+3H^2 q_t (c_t^2-3)\}]}
{2H^2 (3q_t-\Mpl^2)^2}\,,\qquad 
G_{33}=\frac{a^2 c_m^2}{2(\rho+P)}\,.
\ea
To derive these coefficients, we have absorbed $k^2$-dependent 
terms present in ${\bm B}$ into the components of ${\bm G}$ and 
used the relation $C_1=3HC_3-\eta \dot{\phi}$, and 
\be
\dot{C}_3 = C_6+C_3 \left( H+\frac{2\dot{H}}{H} \right)
-\eta \dot{\phi}\,,\qquad
\dot{C}_4 = C_4 \left( H+\frac{2\dot{H}}{H} \right)+C_5\,,
\qquad
\dot{q}_t = H q_t (c_t^2-1)
+\left( H+\frac{\dot{H}}{H} \right) (q_t-\Mpl^2)\,.
\ee
The scalar ghosts are absent under the following three conditions 
\ba
& &
K_{33}=\frac{a^2}{2(\rho+P)}>0\,,
\label{noghost1} \\
& &
K_{11}K_{22}-K_{12}^2
=\frac{3C_4^2\,\eta q_t }{2 (3q_t-\Mpl^2)^2}>0\,,\\
& &
{\rm det}\,{\bm K}=
\frac{3 C_4^2\,\eta q_t a^2}{4(\rho+P) 
(3q_t-\Mpl^2)^2}>0\,.
\label{noghost3} 
\ea
Under the no-ghost condition $q_t>0$ of tensor perturbations, 
inequalities (\ref{noghost1})-(\ref{noghost3}) are satisfied for 
\ba
\rho+P &>& 0\,,\\
\eta &>&0\,.
\ea

In the limit of large $k$, dominant contributions to the second-order 
action (\ref{Ss2}) arise from ${\bm K}$ and ${\bm G}$.
Then, the dispersion relation can be expressed in the form 
\be
{\rm det} \left( c_s^2 {\bm K}-{\bm G} \right)=0\,,
\label{deteq}
\ee
where $c_s$ is the scalar propagation speed.
Solving Eq.~(\ref{deteq}) for $c_s^2$, we obtain the 
following three solutions
\ba
c_{s1}^2 &=& 1\,,\label{c1}\\
c_{s2}^2 &=& 
-\frac{\eta \dot{\phi}^2+\rho+P+3q_t 
[(c_t^2-3)H^2-2\dot{H}]}{6H^2 q_t}\,,\label{cs2} \\
c_{s3}^2 &=& c_m^2\,,\label{c3}
\ea
which correspond to the squared propagation speeds 
of $\delta \phi$, $\delta \chi$, and $\delta \rho$, respectively. 
The scalar perturbation $\delta \phi$ has a luminal propagation 
speed, so it satisfies the Laplacian stability condition. 
For $c_m^2>0$, the matter perturbation $\delta \rho$ is free from 
Laplacian instability.
On using the background Eq.~(\ref{back2}), the sound speed 
squared (\ref{cs2}) can be 
expressed as\footnote{If we eliminate $c_t^2$  
by using Eq.~(\ref{back2}), we can express 
Eq.~(\ref{cs2}) in the form
\be
c_{s2}^2=1+\frac{2\dot{H}}{H^2}+\frac{\eta \dot{\phi}^2+\rho+P}
{3q_t H^2}\,.
\label{csdef}
\ee
{}From this expression, it seems that the existence 
of the last term can lead to 
$c_{s2}^2>0$ even in the decelerating Universe. 
In the absence of matter ($\rho=0=P$), this possibility was 
suggested in Ref.~\cite{DeFelice:2009wp}. 
Eliminating $q_t$ instead of $c_t^2$ from Eq.~(\ref{cs2}), 
it is clear that this possibility is forbidden even 
in the presence of matter.
}
\be
c_{s2}^2=\frac{1}{3} \left( 4-c_t^2+\frac{4\dot{H}}{H^2} 
\right)=-\frac{1}{3} \left( 2+c_t^2+6 w_{\rm eff} 
\right)\,,
\label{cs2f}
\ee
where $w_{\rm eff}$ is the effective equation of state 
defined by Eq.~(\ref{weff}).
The Laplacian stability of $\delta \chi$ is ensured 
for $c_{s2}^2>0$, i.e.,  
\be
w_{\rm eff}<-\frac{1}{6} \left( 2+c_t^2 \right)\,.
\label{weffcon}
\ee
Since we need the condition $c_t^2>0$ for the absence of
Laplacian instability in the tensor sector, 
$w_{\rm eff}$ must be in the range $w_{\rm eff}<-1/3$. 
This translates to the condition 
$\dot{H}+H^2=\ddot{a}/a>0$, so the Laplacian stability 
of $\delta \chi$ requires that the Universe is accelerating. 
In decelerating cosmological epochs, the condition 
(\ref{weffcon}) is always violated for $c_t^2>0$.
During the radiation-dominated ($w_{\rm eff}=1/3$) and 
matter-dominated ($w_{\rm eff}=0$) eras, 
we have $c_{s2}^2=-(4+c_t^2)/3$ and 
$c_{s2}^2=-(2+c_t^2)/3$, respectively, 
which are both negative for $c_t^2>0$.

We thus showed that, for scalar-GB couplings $f(\phi,\G)$ 
containing nonlinear functions of $\G$, $\delta \chi$ 
is prone to the Laplacian instability during the radiation 
and matter eras. Hence nonlinear functions 
of $\G$ should not be present in decelerating cosmological 
epochs. Even if $c_{s2}^2$ is positive in the inflationary 
epoch, $c_{s2}^2$ changes its sign during the transition to 
a reheating epoch (in which $w_{\rm eff} \simeq 0$ for 
a standard reheating scenario). 
During the epoch of late-time cosmic acceleration, 
$c_{s2}^2$ can be positive, but it changes the sign 
as we go back to the matter era.
Since $\delta \chi$ is coupled to $\delta \phi$ and 
$\delta \rho$, the instability of $\delta \chi$ leads to 
the growth of $\delta \phi$ and $\delta \rho$
for perturbations deep inside the Hubble radius. 
This violates the successful background evolution 
during the decelerating cosmological epochs.

The squared propagation speeds (\ref{c1})-(\ref{c3}) 
have been derived by choosing the flat gauge (\ref{flat}), 
but they are independent of the gauge choices. 
Indeed, we will show in Appendix A that the same values of 
$c_{s1}^2$, $c_{s2}^2$, and $c_{s3}^2$ can be obtained 
by choosing the unitary gauge. 
We also note that the scalar propagation speed squared (\ref{cs0})
in $\xi(\phi)\G$ gravity is not equivalent to the value (\ref{c1}).
As we observe in Eq.~(\ref{cs0}), the propagation of $\phi$ is 
affected by the coupling $\xi(\phi)$ with the linear GB term $\G$. 
In $f(\phi,{\cal G})$ theory with $f_{,\G \G} \neq 0$, the new scalar field 
$\chi$ plays a role of the dynamical DOF arising from the nonlinear GB term. 
In this latter case, the propagation of the other field $\phi$ does not practically acquire 
the effect of a coupling with the GB term and hence 
$c_{s1}$ reduces to the luminal value.

\subsection{$f(\G)$ gravity}

Finally, we also study the stability of scalar perturbations
in $f(\G)$ gravity given by the action 
\be
{\cal S}=\int {\rm d}^4 x \sqrt{-g} \left[ 
\frac{\Mpl^2}{2}R+f(\G) \right]+{\cal S}_m (g_{\mu \nu}, \Psi_m)\,.
\label{actionfG}
\ee
In this case, there is no scalar field $\phi$ coupled to the GB term. 
The action (\ref{actionfG}) is equivalent to Eq.~(\ref{action2}) 
with $\phi=0$, $X=0$, $V(\phi)=0$, 
$U=-f(\chi)+\chi \xi(\chi)$, and $\xi=f_{,\chi}(\chi)$. 
As shown in Ref.~\cite{KYY}, this theory 
belongs to a subclass of Horndeski theories with 
one scalar DOF $\chi$ besides a matter fluid. 

In $f(\G)$ gravity, the second-order action of scalar perturbations is 
obtained by setting $\phi$, $\delta \phi$, and their derivatives 0 in 
Eqs.~(\ref{LF}) and (\ref{Lze}). 
We choose the flat gauge (\ref{flat}) and eliminate $\alpha$, $B$, 
$v$ from the action by using Eqs.~(\ref{pereq1})-(\ref{pereq3}). 
Then, the second-order scalar action reduces to the form (\ref{Ss2}) 
with $2 \times 2$ matrices 
${\bm K}$, ${\bm G}$, ${\bm M}$, ${\bm B}$ and
two dynamical perturbations
\be
\vec{\mathcal{X}}^{t}=\left( \delta \chi, 
\delta \rho/k \right) \,.
\ee
In the small-scale limit, nonvanishing components of 
${\bm K}$ and ${\bm G}$ are given by 
\ba
& &
K_{11}=\frac{3C_4^2 q_t}{(3q_t-\Mpl^2)^2}\,,\qquad 
K_{22}=\frac{a^2}{2(\rho+P)}\,,\\
& &
G_{11}=-\frac{C_4^2[\rho+P+3q_t 
[(c_t^2-3)H^2-2\dot{H}]}
{2H^2(3q_t-\Mpl^2)^2}\,,\qquad 
G_{22}=\frac{a^2 c_m^2}{2(\rho+P)}\,.
\ea
The no-ghost conditions correspond to $K_{11}>0$ and $K_{22}>0$, 
which are satisfied for $q_t>0$ and $\rho+P>0$. 
The propagation speed squared for $\delta \chi$ is 
\be
c_{s1}^2=\frac{G_{11}}{K_{11}}=
-\frac{\rho+P+3q_t 
[(c_t^2-3)H^2-2\dot{H}]}{6H^2 q_t}=
-\frac{1}{3} \left( 2+c_t^2+6 w_{\rm eff} 
\right)\,,
\label{cchi}
\ee
where, in the last equality, we used the background 
Eq.~(\ref{back2}) with $\dot{\phi}=0$. 
The other matter propagation speed squared 
is given by $c_{s2}^2=G_{22}/K_{22}=c_m^2$.
Since the last expression of Eq.~(\ref{cchi}) is of the same form 
as Eq.~(\ref{cs2f}), the Laplacian instability of $\delta \chi$ is 
present in decelerating cosmological epochs. 
In Ref.~\cite{DeFelice:2009rw}, violent growth of matter 
perturbations was found during the radiation and matter eras 
for $f(\G)$ models of late-time cosmic acceleration. 
This is attributed to the Laplacian instability of $\delta \chi$ 
coupled to $\delta \rho$, which inevitably occurs for 
nonlinear functions of $f(\G)$.

\section{Conclusions}
\label{consec}

In this letter, we studied the stability of cosmological perturbations 
on the spatially flat FLRW background in scalar-GB theories given 
by the action (\ref{action}). 
Provided that $f_{,\G \G} \neq 0$, the action (\ref{action}) is 
equivalent to (\ref{action2}) with a new scalar DOF $\chi$ 
arising from nonlinear GB terms. 
Theories with $f_{,\G \G}=0$ correspond to a linear GB term 
coupled to a scalar field $\phi$ of the form $\xi(\phi)\G$, 
which belongs to a subclass of Horndeski theories. 
To make a comparison with the scalar-GB coupling $f(\phi,\G)$ 
containing nonlinear functions of $\G$, we first revisited
stabilities of cosmological perturbations in $\xi(\phi)\G$ 
gravity in Sec.~\ref{Fphisec}. 
In this latter theory, provided that the scalar-GB coupling is 
subdominant to the background equations of motion, 
the stability conditions of tensor and scalar perturbations 
can be consistently satisfied.

In Sec.~\ref{backsec}, we derived the background equations and 
stability conditions of tensor perturbations for the scalar-GB 
coupling $f(\phi,\G)$ with $f_{,\G \G} \neq 0$. 
Besides a canonical scalar field $\phi$ with the kinetic term $\eta X$ 
and the potential $V(\phi)$, 
we incorporate a perfect fluid given by 
the Schutz-Sorkin action (\ref{Sm}). 
The absence of ghosts and Laplacian instabilities requires that 
the quantities $q_t$ and $c_t^2$ defined by Eqs.~(\ref{qt}) 
and (\ref{ct}) are both positive. 
In terms of $q_t$ and $c_t^2$, the background equations of motion 
in the gravitational sector can be expressed in a simple manner as 
Eqs.~(\ref{back1}) and (\ref{back2}), where the latter is used 
to simplify a scalar sound speed later. 

In Sec.~\ref{scalarsec}, we expanded the action in $f(\phi,\G)$ 
gravity with $f_{,\G \G} \neq 0$ up to quadratic order 
in scalar perturbations. After eliminating nondynamical 
variables $\alpha$, $B$, and $v$, the second-order 
action is of the form (\ref{Ss2}) with three dynamical 
perturbations (\ref{chit}).
With the no-ghost condition $q_t>0$ of tensor perturbations, 
the scalar ghosts are absent for $\eta>0$ and $\rho+P>0$. 
The sound speeds of perturbations $\delta \phi$ and 
$\delta \rho$ have the standard values $1$ and $c_m$, 
respectively. However, the squared propagation speed 
of $\delta \chi$, which arises from nonlinear GB functions 
in $f$, has a nontrivial value $c_{s2}^2=-(2+c_t^2+6 w_{\rm eff})/3$. 
Since the positivity of $c_{s2}^2$ requires that $w_{\rm eff}<-(2+c_t^2)/6$, 
we have $w_{\rm eff}<-1/3$ under the absence of Laplacian instability 
in the tensor sector ($c_t^2>0$). 
This means that the scalar perturbation associated with nonlinearities 
of the GB term is subject to Laplacian instability during decelerating 
cosmological epochs including radiation and matter eras. 
The same property also holds for $f(\G)$ gravity with 
$f_{,\G \G} \neq 0$. 

We thus showed that a canonical scalar field $\phi$ coupled 
to a nonlinear GB term does not modify the property of 
negative values of $c_{s2}^2$ in the decelerating Universe. 
During inflation or the epoch of late-time cosmic acceleration, 
it is possible to avoid Laplacian instability of the perturbation 
$\delta \chi$ in $f(\phi,\G)$ gravity with $f_{,\G \G} \neq 0$. 
However, in the subsequent reheating period after inflation 
or in the preceding matter era before dark energy dominance, the 
Laplacian instability inevitably emerges to violate the successful 
background cosmological evolution. 
We have shown this for a canonical scalar field $\phi$, but it may be interesting to see whether the same property 
persists for the scalar field $\phi$ arising in Horndeski theories 
and its extensions like DHOST theories \cite{BenAchour:2016cay,Crisostomi:2016czh}. 
While we focused on the analysis on the FLRW background, it will be 
also of interest to study whether some instabilities are present 
for perturbations on a static and spherically symmetric 
background in $f(\phi,\G)$ gravity with $f_{,\G \G} \neq 0$. 
The latter is important for the construction of 
stable hairy black hole or neutron star solutions in theories 
beyond the scalar-GB coupling $\xi(\phi)\G$. 
These issues are left for future works.

\section*{Acknowledgements}

ST is supported by the Grant-in-Aid for Scientific Research 
Fund of the JSPS Nos.~19K03854 and 22K03642.

\appendix

\section{Stability conditions in $f(\phi,\G)$ gravity in unitary gauge}
\label{unitary}

In this Appendix, we derive stability conditions of scalar perturbations in 
$f(\phi,\G)$ gravity by choosing the unitary gauge 
\be
\delta \phi=0\,.
\ee
Then, the gauge-invariant perturbations
${\cal R}=\zeta-H\delta \phi/\dot{\phi}$,  
$\delta \chi_{\rm u}=\delta \chi-\dot{\chi}\delta \phi/\dot{\phi}$, and 
$\delta \rho_{\rm u}=\delta \rho-\dot{\rho}\,\delta \phi/\dot{\phi}$ 
reduce, respectively, to $\zeta$, $\delta \chi$, and $\delta \rho$. 
After the elimination of nondynamical variables $\alpha$, $B$, $v$ from 
Eq.~(\ref{Ss}), the second-order action reduces to the form (\ref{Ss2}) 
with the dynamical perturbations 
\be
\vec{\mathcal{X}}^{t}=\left(\zeta, \delta \chi, 
\delta \rho/k \right) \,,
\ee
where nonvanishing matrix components of ${\bm K}$ and ${\bm G}$ are 
$K_{11}, K_{22}, K_{12}=K_{21}, K_{33}$ and 
$G_{11}, G_{22}, G_{12}=G_{21}, G_{33}$.
In the short-wavelength limit, the ghosts are absent for
\ba
& &
K_{33}=\frac{a^2}{2(\rho+P)}>0\,,
\label{noghost1u} \\
& &
K_{11}K_{22}-K_{12}^2
=\frac{3C_4^2\,\eta q_t \dot{\phi}^2}
{2H^2(3q_t-\Mpl^2)^2}>0\,,\\
& &
{\rm det}\,{\bm K}=
\frac{3 C_4^2\,\eta q_t a^2 \dot{\phi}^2}{4(\rho+P) 
H^2 (3q_t-\Mpl^2)^2}>0\,.
\label{noghost3u} 
\ea
Under the tensor no-ghost condition $q_t>0$,
inequalities (\ref{noghost1u})-(\ref{noghost3u}) are satisfied 
for $\rho+P>0$ and $\eta>0$. 
These conditions are the same as those derived by choosing 
the flat gauge. 

The scalar propagation speed squared $c_s^2$ can be derived 
by solving the dispersion relation (\ref{deteq}). 
On using the background Eq.~(\ref{back2}), 
we obtain the three values of $c_s^2$ exactly the same as 
Eqs.~(\ref{c1})-(\ref{c3}). 
Thus, the propagation speeds in the small-scale limit are 
independent of the gauge choices. 
In $f(\G)$ gravity, we also obtain the same scalar propagation 
speeds as those derived in the flat gauge.

\bibliographystyle{mybibstyle}
\bibliography{bib}

\end{document}